\documentclass[aps,pra,superscriptaddress,notitlepage,groupaddress,showpacs,nofootinbib,twocolumn]{revtex4-1}
\usepackage{times,graphics,graphicx,epsf,epsfig,color,xcolor,amsfonts,amssymb,amsmath,bbm,bm,dsfont,hyperref}
\newtheorem{lemma}{Lemma}
\newtheorem{theorem}{Theorem}
\hypersetup{colorlinks,linkcolor=blue,citecolor=blue,urlcolor=blue}

\let\oldsqrt\sqrt
\def\sqrt{\mathpalette\DHLhksqrt}
\def\DHLhksqrt#1#2{%
\setbox0=\hbox{$#1\oldsqrt{#2\,}$}\dimen0=\ht0
\advance\dimen0-0.2\ht0
\setbox2=\hbox{\vrule height\ht0 depth -\dimen0}%
{\box0\lower0.4pt\box2}}

\newcommand{\average}[1]{\left\langle#1\right\rangle}
\newcommand{\ignore}[1]{}

\DeclareFontFamily{OT1}{pzc}{}
\DeclareFontShape{OT1}{pzc}{m}{it}%
              {<-> s * [1.25] pzcmi7t}{}
\DeclareMathAlphabet{\mathpzc}{OT1}{pzc}%
                                 {m}{it}

\begin{document}

\title{Bound on the rate of entropy change in open quantum systems}

\author{F. Bakhshinezhad}
\affiliation{Department of Physics, Sharif University of Technology, Tehran 14588, Iran}
\affiliation{Institute for Quantum Optics and Quantum Information - IQOQI Vienna, Austrian Academy of Sciences, Boltzmanngasse 3, 1090 Vienna, Austria}

\author{A. T. Rezakhani}
\affiliation{Department of Physics, Sharif University of Technology, Tehran 14588, Iran}

\begin{abstract}
We study the temporal rate of variations of the von Neumann entropy in an open quantum system which interacts with a bath. We show that for almost all initial states of the bath and the system, the time-average of the rate of entropy change is bounded by a function which depends on various properties of the system and environment, and is mostly relatively small. This result holds true under fairly general conditions in almost any arbitrary quantum system.
\end{abstract}
\maketitle
 
 \section{Introduction}
\label{sec:int}

An open quantum system inevitably interacts with its environment (or ``bath") \cite{ref-Bruer}. Such interactions may typically lead to loss of quantum information or quantum features (such as coherence and correlations) within the system, and in turn affect its dynamics \cite{ref-Zurek}. This highlights that studying how quantum information vary in a generic quantum system is important. The quantity that captures this behavior is entropy \cite{ref-Nielsen, ref-Bravyi}, which also plays a principal role in describing relevant statistical mechanics of the system \cite{ref-Eisert,ref-Gogolin,ref-us}. Quantifying variations of entropy can be useful in understanding dynamical behavior of the system \cite{ref-Guzik PRL, ref-Wehner PRL,ref-Lieb, ref-Acoleyen,ref-Short T, ref-Short NJP, ref-Reimann PRL, ref-Malabarba, ref-Wilming,ref-Das} and whether and how it approaches thermal equilibrium \cite{ref-Popescu PRE, ref-Popescu NJP}.
 
Noting that the calculation of entropy in open quantum systems is generally a formidable task, obtaining bounds on the rate of entropy change becomes important and useful per se \cite{ ref-Bravyi, ref-Wehner PRL, ref-Lieb, ref-Acoleyen}. Here, built up on earlier literature, we obtain bounds on the finite-time temporal average of the rate of entropy change. We also investigate how on average initial states of the system-bath affects this time-average rate. We show that for typical cases of a sufficiently large bath the average of the rate of entropy change becomes sufficiently small for almost all preparations of the system-bath. 
 
The outline of this short paper is as follows. In Sec. \ref{sec:pre}, we briefly remind some preliminaries. In Sec. \ref{sec:results}, we state our main results. We summarize our findings in Sec.~\ref{sec:summary}.

\section{Preliminaries}
\label{sec:pre}

Consider a closed composite quantum system $\mathsf{SB}$ comprised of two parts, ``system" $\mathsf{S}$ and ``bath" $\mathsf{B}$, with the Hilbert space $\mathpzc{H}_{~\mathsf{SB}}= \mathpzc{H}_{~\mathsf{S}} \otimes \mathpzc{H}_{~\mathsf{B}}$, where $\dim (\mathpzc{H}_{~\mathsf{S,B}}) = d_{\mathsf{S,B}}<\infty$. In typical scenarios, there may be some constraints on the dynamics, such as existence of specific conserved observables, which can be enforced by restricting the allowed states to a certain subspace $\mathpzc{H}_{\,\mathsf{R}} \subseteq \mathpzc{H}_{~\mathsf{S}} \otimes \mathpzc{H}_{~\mathsf{B}}$ \cite{ref-Popescu Nature}.

\begin{figure}[bp]
\includegraphics[scale=0.25]{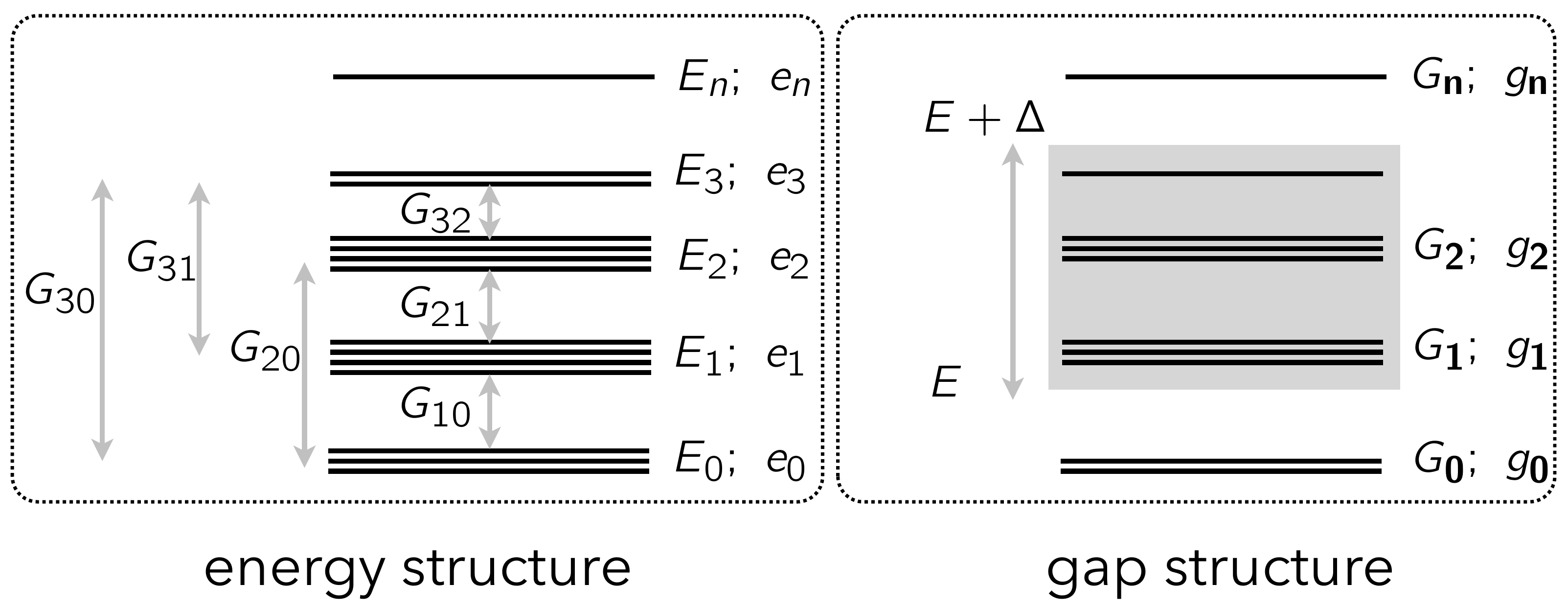}
\caption{Energy and gap structures. Each energy eigenvalue $E_n$ has the degeneracy $e_n$ ($n\in\{0,1,\ldots,D_E -1\}$), and each gap value $G_{\mathbf{n}}$ has the degeneracy $g_{\mathbf{n}}$. The quantity $N(\Delta)$ is obtained by counting the number of gaps in the interval $[E,E+\Delta)$ and sweeping over all $E$s in the gap structure to find an $E$ for which this number is maximum for a given $\Delta$.}
\label{fig:1}
\end{figure}

Let the system and the bath evolve with the (time-independent) Hamiltonian $\mathbbmss{H}_{\mathsf{SB}} = \mathbbmss{H}_{\mathsf{S}} + \mathbbmss{H}_{\mathsf{B}} + \mathbbmss{H}_{\mathsf{int}}$, where $ \mathbbmss{H}_{\mathsf{S}}$, $ \mathbbmss{H}_{\mathsf{B}}$, and $\mathbbmss{H}_{\mathsf{int}}$ are, respectively, the system, bath, and interaction Hamiltonians. Assume the spectral decomposition 
\begin{equation}
\mathbbmss{H}_{\mathsf{SB}} = \sum_{n=0}^{D_E-1} E_n \mathpzc{P}_n,
\end{equation}
where $E_n$s are \textit{distinct} eigenvalues (obviously $D_{E}\leqslant d_{\mathsf{S}}d_{\mathsf{B}}$) and $\mathpzc{P}_n=\sum_{\alpha =1}^{e_n}|n,\alpha\rangle\langle n,\alpha|$ 
is the projection onto the eigensubspace corresponding to the ($e_n$-fold degenerate) eigenvalue $E_n$, with the orthonormality property $\mathpzc{P}_{n}\mathpzc{P}_{n'}=\delta_{nn'}\mathpzc{P}_{n}$ and the completeness property $\sum_{n}\mathpzc{P}_{n}=\mathbbmss{I}_{\mathsf{SB}}$ (the identity operator). We also define $D_{g}=\max_{n}e_{n}$. A Hamiltonian is called \textit{nonresonant} when its energy gaps $G_{ij} \equiv E_i - E_j$ (for $i\neq j$) are nondegenerate. In other words, if we have four eigenvalues $E_i$, $E_j$, $E_k$, and $E_l$ satisfying the equation $E_i - E_j = E_k - E_l$, this should yield either $(i,j)=(k,l)$ or $(i,l)=(j,k)$ \cite{ref-Popescu PRE}. In addition, assume that each gap value $G_{\mathbf{n}}$ has the degeneracy $g_{\mathbf{n}}$, where $\mathbf{n}$ denotes the labels of the gap values. We also denote the largest degeneracy of the energy gaps with $D_G\equiv \max_{\mathbf{n}} g_{\mathbf{n}}$. Density of the energy gaps is captured by the maximum number of the energy gaps $N(\Delta)$ in each energy interval $\Delta > 0$---Fig.~\ref{fig:1}. Note that $D_{G}=\lim_{\Delta \rightarrow 0} N(\Delta)$ \cite{ref-Short T}. 

Assume that the composite system is in a pure state $\varrho_{\mathsf{SB}}(\tau) = | \varphi(\tau)\rangle_{\mathsf{SB}}\langle \varphi(\tau)|$, where $|\varphi(\tau)\rangle_{\mathsf{SB}}=e^{-i\tau \mathbbmss{H}_{\mathsf{SB}}}|\varphi(0)\rangle_{\mathsf{SB}}$ (presuming $\hbar\equiv 1$), and from whence the states of the system and the bath are obtained as $\varrho_{\mathsf{S,B}}(\tau) = \mathrm{Tr}_{\mathsf{B,S}}\big[\varrho_{\mathsf{SB}}(\tau)\big]$. The time-averaged state of the composite system is given by
\begin{align}
\omega_{\mathsf{SB}} \equiv& \langle \varrho_{\mathsf{SB}}(\tau) \rangle_{T\to \infty}= \lim_{T\rightarrow \infty} \frac{1}{T}\int_0^T  \varrho_{\mathsf{SB}}(\tau)\, \mathrm{d}\tau. \label{eq-3}
\end{align}
From this definition, it is evident that $\omega_{\mathsf{SB}}=\sum_{n}\mathpzc{P}_n \varrho_\mathsf{SB}(0) \mathpzc{P}_n$ and $[\omega_{\mathsf{SB}},\mathbbmss{H}_{\mathsf{SB}}]=0$. A relevant quantity, the ``effective dimension" of the state is also defined as \cite{ref-Short T}
\begin{equation}
D_{\mathrm{eff}}\equiv 1/\textstyle{\sum_n} \big(\mathrm{Tr}[\mathpzc{P}_n \,\varrho_{\mathsf{SB}}(0)\mathpzc{P}_{n}]\big)^2.
\label{eq-5}
\end{equation}
Note that $1\leqslant D_{\mathrm{eff}} \leqslant D_{E}$. We also need to remind the definition of the von Neumann entropy $\mathbbmss{S}(\varrho) = -\mathrm{Tr}[\varrho\ln\varrho]$. Table \ref{tab1} summarizes the notations and definition we use throughout this paper.
\begin{table}[tp]
\caption{List of the notations and definitions.}
\begin{ruledtabular}
\begin{tabular}{ll}
$\mathpzc{H}_{\,\mathsf{a}}$ & Hilbert space of system $\mathsf{a}$ ($\in\{\mathsf{S},\mathsf{B},\mathsf{SB},\mathsf{R}\}$)\\
$d_{\mathsf{a}}$ & dimension of $\mathpzc{H}_{\,\mathsf{a}}$\\
$\Pi_{\mathsf{a}}$ & projection on $\mathpzc{H}_{\,\mathsf{a}}$\\
$\varrho_{\mathsf{a}}$ & state of system $\mathsf{a}$ \\
$\{\sigma_{k}\}$ & orthonormal operator basis on $\mathpzc{H}_{\,\mathsf{S}}$ ($\mathrm{Tr}[\sigma^{\dag}_{k}\sigma_{k'}]=\delta_{kk'}$)\\
$E_{n}$ & distinct energy eigenvalues of $\mathbbmss{H}_{\mathsf{SB}}$\\
$e_{n}$ & degeneracy of $E_{n}$ ($\mathrm{Tr}[\mathpzc{P}_{n}]$)\\
$D_{g}$ & $\max_{n} e_{n}$\\
$D_{E}$ & number of distinct $E_{n}$s \\
$\mathpzc{P}_{n}$ & eigenprojection corresponding to $E_{n}$\\
$G_{\mathbf{n}}$ & energy gap $E_{i}-E_{j}$ ($\mathbf{n}\equiv(i,j)$, for $i\neq j$)\\
$g_{\mathbf{n}}$ & degeneracy of the gap $G_{\mathbf{n}}$ \\
$D_{G}$ & $\max_{\mathbf{n}} g_{\mathbf{n}}$ \\
$N(\Delta)$ &  $\max_{E}|\{\mathbf{n};~G_{\mathbf{n}}\in[E,E+\Delta)\}|$ (Fig. \ref{fig:1})\\
$\Delta_{\min}$ & $\min_{\mathbf{n},\mathbf{n}'}\{|G_{\mathbf{n}}-G_{\mathbf{n}'}|;~G_{\mathbf{n}}\neq G_{\mathbf{n}'}\}$\\
$\langle X(\tau) \rangle_{T}$ & $\frac{1}{T}\int_{0}^{T}X(\tau)\,\mathrm{d}\tau$\\
$\omega_{\mathsf{SB}}$ & $\langle \varrho_{\mathsf{SB}}(\tau) \rangle_{T\to\infty}$ \\
$D_{\mathrm{eff}}$ & effective dimension (\ref{eq-5}) \\
$\Vert X\Vert_1$ & trace norm $\mathrm{Tr}[\sqrt{X^{\dag}X}]$\\
$\Vert X\Vert_2$ & Hilbert-Schmidt norm $\sqrt{\mathrm{Tr}[X^{\dag}X]}$\\
$\Vert X\Vert$ & standard operator norm ($\max_{\Vert v\Vert=1}\Vert X|v\rangle\Vert$)\\
$\Vert X\Vert'$ & $\min_{c\in \mathbb{R}}\Vert X-c\mathbbmss{I}\Vert$\\
$h$ & $\Vert \mathbbmss{H}_{\mathsf{S}}+\mathbbmss{H}_{\mathsf{int}} \Vert'$\\
$\mathcal{S}$ & swap operator ($\mathcal{S}(|u\rangle\otimes |v\rangle)=|v\rangle\otimes |u\rangle$)\\
$\mathbbmss{S}(\varrho)$ & von Neumann entropy ($-\mathrm{Tr}[\varrho \ln \varrho]$)\\
\end{tabular}	
\end{ruledtabular}
\label{tab1}
\end{table}

The following lemma is essential for obtaining our main result:
\begin{lemma} \cite{ref-Short T}
\label{t-small}
For any $\Delta > 0$, any time $T > 0$, and any observable $O$ defined on $\mathpzc{H}_{\,\mathsf{R}}$, we have 
\begin{align}
&\big\langle \big|\mathrm{Tr}[\varrho(\tau)O]- \mathrm{Tr}[\omega O]\big|^2\big\rangle_{T}\leqslant \frac{N(\Delta)\|O\|'^{\,2}}{D_{\mathrm{eff}}}\Big(1+ \frac{8\log_2D_E}{\Delta \,T}\Big).
\label{eq-9}\\
&\big\langle \big|\mathrm{Tr}[\varrho(\tau)O]- \mathrm{Tr}[\omega O]\big|^2\big\rangle_{T}\leqslant \frac{D_G\,\|O\|'^{\,2}}{D_{\mathrm{eff}}}\Big(1+ \frac{8\log_2D_E}{\Delta_{\min}\, T}\Big).
\label{eq-10}
\end{align}
Here $\varrho$ and $\omega$ are shorthands for $\varrho_{\mathsf{SB}}$ and $\omega_{\mathsf{SB}}$, respectively.
\end{lemma}

\section{Main result}
\label{sec:results}

By a straightforward modification of Lemma \ref{t-small} we first prove the following lemma:
\begin{lemma}
\label{lemma-2}
For any $T\geqslant 0$, 
\begin{equation}
\Big\langle \Big\|\frac{\mathrm{d}\varrho_\mathsf{S}(\tau)}{\mathrm{d}\tau}\Big\|_1\Big\rangle_T \leqslant 2 h\sqrt{\frac{D_G\, d_\mathsf{S}^{2}}{D_{\mathrm{eff}}(\omega_{\mathsf{SB}})}\Big(1+ \frac{8\log_2D_E}{\Delta_\mathrm{min}\, T}\Big)}.
\label{eq-State rate}
\end{equation}
\end{lemma}

\textit{Proof}: Our proof follows closely the proof of the main result of Ref. \cite{ref-Popescu NJP}. We write 
\begin{equation}
 \frac{\mathrm{d}\varrho_\mathsf{S}(\tau)}{\mathrm{d}\tau}= \mathrm{Tr}_\mathsf{B}\Big[i[\varrho_\mathsf{SB}(\tau),\mathbbmss{H}_{\mathsf{SB}}]\Big]=\sum_{k=0}^{d_\mathsf{S}^2-1} b_k(\tau)\sigma_k,
 \label{eq-state rate2}
 \end{equation}
in which $\{\sigma_k\}$ is an orthonormal operator basis defined on $\mathpzc{H}_{\,\mathsf{S}}$ \cite{ref-Short T}. After some algebra one can see that 
\begin{align}
 b_k(\tau)  = \mathrm{Tr}_\mathsf{SB}\Big[i[\mathbbmss{H}_{\mathsf{S}}+\mathbbmss{H}_{\mathsf{int}}-c\mathbbmss{I}_\mathsf{SB},\sigma_k \otimes \mathbbmss{I}_\mathsf{B}]\varrho_\mathsf{SB}(\tau)\Big],  \label{eq-state rate4}
\end{align}
where we have used $[\mathbbmss{H}_{\mathsf{B}},\sigma_k \otimes \mathbbmss{I}_\mathsf{B}]=0$, the cyclic property of the trace ($\mathrm{Tr}[[A,B]C]=\mathrm{Tr}[A[B,C]]$), and have added a multiple of the identity operator ($-c\mathbbmss{I}_{\mathsf{SB}}$). Replacing $ O=i[\mathbbmss{H}_{\mathsf{S}}+\mathbbmss{H}_{\mathsf{int}}-c\mathbbmss{I}_\mathsf{SB},\sigma_k \otimes \mathbbmss{I}_\mathsf{B}]$ in Eq.~(\ref{eq-10}) gives
\begin{equation}
 \average{|b_k(\tau)-\average{b_k(\tau)}_{T\to \infty}|^2}_{T} \leqslant \frac{D_G\,\|  O\|'^{\,2}} {D_{\mathrm{eff}}}\Big(1+\frac{8\log_{2} D_E}{\Delta_{\mathrm{min}}\,T}\Big). 
 \label{eq-state rate5}
 \end{equation}
Note that 
 \begin{align}
\average{b_k(\tau)}_{T\to\infty}
&=i\,\mathrm{Tr}_\mathsf{SB}\big[[\omega_\mathsf{SB},\mathbbmss{H}_{\mathsf{SB}}]\sigma_k\otimes \mathbbmss{I}_\mathsf{B}\big]=0,
\label{ave ck}\\
\|O\|'^{\,2}  &\leqslant  4 h^{2} \|\sigma_k \otimes \mathbbmss{I}_\mathsf{B}\|^2 =4 h^{2} \|\sigma_k \|^2,
 \label{eq-state rate6}
 \end{align}
where $h\equiv \min_{c\in \mathbb{R}}\Vert \mathbbmss{H}_{\mathsf{S}}+\mathbbmss{H}_{\mathsf{int}} -c\mathbbmss{I}\Vert$ and in the last line we have used the triangle inequality ($\Vert X+Y\Vert\leqslant \Vert X\Vert + \Vert Y\Vert$) and the submultiplicativity property ($\Vert XY \Vert\leqslant \|X \|\|Y \|$) \cite{ref-Bathia Book}. Hence, Eq.~(\ref{eq-state rate5}) reduces to
\begin{equation}
 \average{|b_k(\tau)|^2}_{T} \leqslant \frac{4D_G\,h^{2}\|\sigma_k\|^2}{ D_{\mathrm{eff}}}\Big(1+\dfrac{8\log_{2} D_E}{\Delta_{\mathrm{min}}\, T}\Big).
\label{eq-ck}
\end{equation}

Now, we employ the above inequality to derive a bound on the speed of the state change as
\begin{align}
\Big\langle\Big\|\frac{\mathrm{d}\varrho_\mathsf{S}(\tau)}{\mathrm{d}\tau}\Big\|_1\Big\rangle_T &\leqslant \sqrt{d_\mathsf{S}} \Big\langle\Big\|\dfrac{d\varrho_\mathsf{S}(\tau)}{\mathrm{d}\tau}\Big\|_2\Big\rangle_T \nonumber\\
&\overset{\mathrm{(\ref{eq-state rate2})}}{ = }  \sqrt{d_\mathsf{S}\sum_{k=0}^{d_\mathsf{S}^2-1} \average{|b_k(\tau)|^2}_T} \nonumber\\
&\overset{\mathrm{(\ref{eq-ck})}}{\leqslant}  2h \sqrt{\dfrac{D_G\,d_\mathsf{S}^{2}}{ D_{\mathrm{eff}}}\Big(1+\dfrac{8\log_2 D_E}{\Delta_{\mathrm{min}}\, T}\Big)}, 
\end{align}
where we have used the properties $\|X\|_1\leqslant \sqrt{\mathrm{rank}(X)}\|X\|_2$ and $\Vert X \Vert\leqslant \Vert X\Vert_{2}$ and the concavity of the square-root function ($\langle \sqrt{X}\rangle \leqslant \sqrt{\langle X\rangle}$) \cite{ref-Bathia Book}. 
 \hfill $\blacksquare$

Note that in the long-time limit $T\to\infty$ and for Hamiltonians with nondegenerate gaps (i.e., $D_{G}=1$), Eq. (\ref{eq-State rate}) becomes 
\begin{equation}
\Big\langle\Big\|\frac{\mathrm{d}\varrho_{\mathsf{S}}\left(\tau\right)}{\mathrm{d}\tau}\Big\|_1\Big\rangle_{T \to \infty} \leqslant 2 h \sqrt{\frac{d_{\mathsf{S}}^2}{D_{\mathrm{eff}}}},
\label{eq-12}
\end{equation}
which is compatible with the result of Ref. \cite{ref-Popescu NJP}.

\begin{theorem}
\label{thm:1}
Consider a quantum system $\mathsf{S}$ coupled to a bath $\mathsf{B}$, which jointly evolve under a time-independent Hamiltonian. Assume $d_{\mathsf{S}}^3 \leqslant d_{\mathsf{B}}$ and take $\eta$ an arbitrary number satisfying $\sqrt{d_{\mathsf{S}} / d_{\mathsf{B}}} \leqslant \sqrt{ d_{\mathsf{S}} /d_{\mathsf{B}}} + \eta \leqslant 1/d_{\mathsf{S}}$. Now if we draw initial states from $\mathpzc{H}_{~\mathsf{SB}}$ uniformly randomly and calculate the rate of entropy change, then the probability for the finite-time average of the entropy rate satisfies the following property:
\begin{equation}
P_{\varphi_{\mathsf{SB}}} \left[ \Big\langle \Big|\frac{\mathrm{d}\mathbbmss{S}(\varrho_{\mathsf{S}} (\tau))}{\mathrm{d}\tau}\Big| \Big\rangle_T   \geqslant \delta \right] \leqslant \varepsilon,
\label{eq-15}
\end{equation}
where 
\begin{align}
\delta=& 2 h \big( \sqrt{d_{\mathsf{S}}/d_{\mathsf{B}}} + \eta \big) \sqrt{\frac{D_G\, d_{\mathsf{S}}^{4}}{D_{\mathrm{eff}}}\left(1+\frac{8 \log_2 D_E}{\Delta_{\mathrm{min}}\, T}\right)},\\
\varepsilon = & 2 e^{-d_{\mathsf{S}} d_{\mathsf{B}} \eta^2/16}.
\end{align}
\end{theorem}

\textit{Proof}: We follow steps similar to Ref. \cite{ref-Wehner PRL}. Note that \cite{ref-Alicki}
\begin{align}
\frac{\mathrm{d}\mathbbmss{S}\left(\varrho_{\mathsf{S}}(\tau)\right)}{\mathrm{d}\tau} &=    
\mathrm{Tr}_{\mathsf{S}}\Big[\Big(\ln \varrho_{\mathsf{S}}(\tau) - \ln\frac{\mathbbmss{I}_{\mathsf{S}}}{d_{\mathsf{S}}}\Big) \frac{\mathrm{d}\varrho_{\mathsf{S}}(\tau)}{\mathrm{d}\tau}\Big],
\label{eq-16}
\end{align} 
which yields
\begin{align}
\Big|\frac{\mathrm{d}\mathbbmss{S}\left(\varrho_{\mathsf{S}}(\tau)\right)}{\mathrm{d}\tau}\Big| &\leqslant \Big\|\ln \varrho_{\mathsf{S}}(\tau) - \ln\frac{\mathbbmss{I}_{\mathsf{S}}}{d_{\mathsf{S}}} \Big\| \Big\|\frac{\mathrm{d}\varrho_{\mathsf{S}}(\tau)}{\mathrm{d}\tau}\Big\|_1,
\label{eq-18}
\end{align}
where we have used the inequality $|\mathrm{Tr}[XY]| \leqslant \|X\|_1 \|Y\|$ \cite{ref-Bathia Book}. Now, let $\left\{r_i\right\}_{i=0}^{d_{\mathsf{S}}-1} $ denote the eigenvalues of $\varrho_{\mathsf{S}}(\tau)$; hence, Eq. (\ref{eq-18}) reduces to
\begin{equation}
\Big|\frac{\mathrm{d}\mathbbmss{S}\left(\varrho_{\mathsf{S}}(\tau)\right)}{\mathrm{d}\tau}\Big|\leqslant \Big\|\frac{\mathrm{d}\varrho_{\mathsf{S}}(\tau)}{\mathrm{d}\tau}\Big\|_1  . \max_{i} |\ln (r_i\, d_{\mathsf{S}})|.
\label{eq-20}
\end{equation}
Now we recall two results from Ref. \cite{ref-Wehner PRL}: (i) for $d_{\mathsf{S}}\Vert\varrho_{\mathsf{S}}(\tau) - \mathbbmss{I}_{\mathsf{S}}/d_{\mathsf{S}} \Vert_1 \leqslant 1$ and $d_{\mathsf{S}}\geqslant 2$, we have
\begin{equation} 
\max_{i}\left|\ln(r_i\, d_{\mathsf{S}})\right|\leqslant d_{\mathsf{S}}\Vert\varrho_{\mathsf{S}}(\tau) - \mathbbmss{I}_{\mathsf{S}}/d_{\mathsf{S}} \Vert_1.
\label{eq-21}
\end{equation}
(ii) If we choose initial states $|\varphi(0)\rangle_{\mathsf{SB}}$ (shortly $\varphi_{\mathsf{SB}}$) of the composite system \textit{uniformly} randomly from $\mathpzc{H}_{~\mathsf{SB}}$ and then calculate $\Vert\varrho_{\mathsf{S}}(\tau) - \mathbbmss{I}_{\mathsf{S}}/d_{\mathsf{S}} \Vert_1$, we obtain 
\begin{equation}
P_{\varphi_{\mathsf{SB}}}\left[ \Vert\varrho_{\mathsf{S}}(\tau) - \mathbbmss{I}_{\mathsf{S}}/d_{\mathsf{S}} \Vert_1\geqslant \sqrt{ d_{\mathsf{S}} /d_{\mathsf{B}}} + \eta \right] \leqslant 2 e^{-d_{\mathsf{S}}d_{\mathsf{B}} \eta^2/16}.
\label{eq-23}
\end{equation}
Combining all pieces now yields the desired result. \hfill$\blacksquare$

From this theorem it is evident that for a sufficiently large bath ($d_{\mathsf{B}}\gg d^{3}_{\mathsf{S}}$) one can make $\varepsilon$ sufficiently small. That is, for such systems the rate of entropy change is \textit{almost always} (i.e., with a probability $\geqslant 1-\varepsilon$, for $\varepsilon\ll1$) negligibly small. This result is compatible with the fact that sufficiently small subsystems of a large system in a pure state look relatively similar to the maximally mixed state \cite{ref-Popescu Nature}, because such states do not change appreciably.   

Next, we calculate the bound for all initial states and compute the average of this quantity over all possible pure states. To do so, we need to calculate the ensemble average of relation (\ref{eq-15}). The only parameter on the right-hand side of this inequality which depends on the initial state is $\sqrt{1/D_{\mathrm{eff}}}$. We use the convexity property of the square-root function ($\langle \sqrt{{1}/{D_{\mathrm{eff}}}}\rangle_{\varphi_{\mathsf{SB}}}\leqslant \sqrt{\langle {{1}/{D_{\mathrm{eff}}}}\rangle_{\varphi_{\mathsf{SB}}}}$). Note that 
\begin{align}
\Big\langle \frac{1}{D_{\mathrm{eff}}}\Big\rangle_{\varphi_{\mathsf{SB}}} =\sum_{n}\mathrm{Tr}\left[\mathpzc{P}_n\otimes \mathpzc{P}_n \average{|\varphi\rangle_{\mathsf{SB}} \langle \varphi| \otimes |\varphi\rangle_{\mathsf{SB}} \langle \varphi|}_{\varphi_{\mathsf{SB}}}\right], \nonumber
\end{align}
where we have used the identity $\mathrm{Tr}[X]\,\mathrm{Tr}[Y]=\mathrm{Tr}[X\otimes Y]$. Now we employ the relation
\begin{equation}
\big\langle |\varphi\rangle_{\mathsf{SB}} \langle \varphi| \otimes |\varphi\rangle_{\mathsf{SB}} \langle \varphi|\big\rangle_{\varphi_{\mathsf{SB}}}=\frac{\Pi_{\,\mathsf{R}}\otimes\Pi_{\,\mathsf{R}}(\mathbbmss{I}+\mathcal{S})}{d_{\,\mathsf{R}}(d_{\,\mathsf{R}}+1)},
\label{eq-ens. ave. state}
\end{equation}
in which $\Pi_{\,\mathsf{R}}$ and $\mathcal{S}$ denote the projector onto $\mathpzc{H}_{\,\mathsf{R}}$ and the swap operator ($\mathcal{S}(X\otimes Y)\mathcal{S}=Y\otimes X$), respectively \cite{ref-Popescu PRE}; whence 
\begin{align}
\Big\langle \frac{1}{D_{\mathrm{eff}}}\Big\rangle_{\varphi_{\mathsf{SB}}} &= \sum_{n=0}^{D_{E}-1}\mathrm{Tr}\left[\mathpzc{P}_n\otimes \mathpzc{P}_n\frac{\Pi_{\,\mathsf{R}}\otimes\Pi_{\,\mathsf{R}}(\mathbbmss{I}+\mathcal{S})}{d_{\,\mathsf{R}}(d_{\,\mathsf{R}}+1)} \right]\nonumber\\
&=\frac{2}{d_{\,\mathsf{R}}(d_{\,\mathsf{R}}+1)}\sum_{n}\mathrm{Tr}[\mathpzc{P}_n\otimes \mathpzc{P}_n (\Pi_{\,\mathsf{R}}\otimes\Pi_{\,\mathsf{R}})]\nonumber\\
&\leqslant \frac{2}{d_{\,\mathsf{R}}(d_{\,\mathsf{R}}+1)}\sum_{n}\mathrm{Tr}[\mathpzc{P}_n\otimes \mathpzc{P}_n]\nonumber\\
& = \frac{2\sum_{n}~e_n^2}{d_{\,\mathsf{R}}(d_{\,\mathsf{R}}+1)}
\leqslant \frac{2D_g^2 \,D_E}{d_{\,\mathsf{R}}(d_{\,\mathsf{R}}+1)}.
\label{eq-ens. ave. eff2}
\end{align}

Using Eqs.~(\ref{eq-15}) and (\ref{eq-ens. ave. eff2}) in the limit of large bath dimension, we can see that the time and initial state averages of the rate of entropy change is bounded by 
\begin{align}
&\Big\langle \Big|\frac{\mathrm{d}\mathbbmss{S}(\varrho_{\mathsf{S}} (\tau))}{\mathrm{d}\tau}\Big| \Big\rangle_{T,\varphi_{\mathsf{SB}}}   \leqslant  h \big( \sqrt{d_{\mathsf{S}}/d_{\mathsf{B}}} + \eta \big)\times\nonumber\\
&\, \sqrt{\frac{8 D_G\,D_E\,D_g^2\,d_{\mathsf{S}}^4}{d_\mathsf{R}(d_\mathsf{R}+1)} \left(1+\frac{8 \log_2 D_E}{\Delta_{\min}\, T}\right)}.
\label{eq-25}
\end{align}
Now by assuming $T\rightarrow \infty$, $d_\mathsf{R} \approx d_{\mathsf{B}}d_{\mathsf{S}}$, and a nondegenerate and nonresonant Hamiltonian ($D_{g}=D_{G}=1$), we obtain
\begin{equation}
\Big\langle\Big|\frac{\mathrm{d}\mathbbmss{S}(\varrho_{\mathsf{S}} (\tau))}{\mathrm{d}\tau}\Big| \Big\rangle_{T\to\infty,\varphi_{\mathsf{SB}}}   \leqslant    \sqrt{8} h\frac{d_{\mathsf{S}}^2}{d_{\mathsf{B}}}.
\label{eq-26}
\end{equation} 

\section{Summary}
\label{sec:summary}

We have obtained an upper bound on the rate of entropy change for open systems interacting with a bath. In particular, we have shown that if an initial state is chosen uniformly, with a considerable probability the rate of entropy change of the system can be significantly small at any time, if the following conditions are met: (i) the energy gaps between \textit{distinct} energy levels have relatively small degeneracy, (ii) the dimension of the Hilbert space of the system is sufficiently small compared to the dimension of the Hilbert space of the bath, and (iii) the system and the interaction Hamiltonians do not have a relatively wide spectrum. The bound we have obtained depends on the initial state of the composite system through the effective dimension. But we have shown that if the initial state is spread over many different eigenvectors of the total Hamiltonian (which in turn implies a higher effective dimension), the average rate of entropy change becomes relatively small. Our results hold for almost all systems and imply that the rate of information loss in such systems (small systems coupled to a relatively larger bath, both with high dimensions) becomes small. Because the number of initial configurations of the composite system which violate our bound is negligible, we have taken average over all initial states, which has yielded a bound which is (almost) independent of the initial state of the composite system and its effective dimension.

\textit{Acknowledgments.}---Initial inputs of P. Asadi and S. A. Seif Tabrizi are acknowledged. This work was partially supported by Sharif University of Technology's Office of Vice President for Research (through Contract QA960512). F.B. also acknowledges support from the Ministry of Science Research and Technology of Iran and the Austrian Science Fund (FWF) through the START project Y879-N27.



\end{document}